\documentclass[aps,amsmath,amssymb,preprint,citeautoscript]{revtex4}

\usepackage{graphicx}

\usepackage{amsmath}

\begin{document}

\title{Phase-diagram of two-color lattice QCD in the chiral limit}
\author{Shailesh Chandrasekharan and Fu-Jiun Jiang}
\affiliation{Department of Physics, Box 90305, Duke University,
Durham, North Carolina 27708.}

\begin{abstract}

We study thermodynamics of strongly coupled lattice QCD with two colors 
of massless staggered fermions as a function of the baryon chemical 
potential $\mu$ in $3+1$ dimensions using a new cluster
algorithm. We find evidence that the model undergoes a weak first order 
phase transition at $\mu=0$ which becomes second order at a finite $\mu$. 
Symmetry considerations suggest that the universality class of these 
phase transitions should be governed by a $O(N)\times O(2)$ field theory 
with collinear order, with $N=3$ at $\mu=0$ and $N=2$ at $\mu \neq 0$. 
The universality class of the second order phase transition at $\mu\neq 0$ 
appears to be governed by the decoupled $XY$ fixed point present in the 
$O(2)\times O(2)$ field theory. Finally we show that the quantum ($T=0$) 
phase transition as a function of $\mu$ is a second order mean field 
transition.
\end{abstract}

\maketitle

\section{INTRODUCTION}

Understanding the phase diagram of Quantum Chromodynamics (QCD), as a 
function of temperature $(T)$ and baryon chemical potential $(\mu)$ is 
of great interest in the phenomenology of strongly interacting dense 
matter \cite{Rajagopal:2000wf}. There are many excellent reviews on the 
subject and some recent ones can be found in 
\cite{Schafer:2003vz,Rischke:2003mt,Alford:2003eg,
Stephanov:2004wx}. The only first principles approach to the subject is 
based on the lattice formulation of QCD in which one computes quantities 
using the Monte Carlo method. Unfortunately, due to a variety of algorithmic 
difficulties this has been difficult to accomplish. At intermediate and
large chemical potentials and small temperatures the numerical methods 
suffer from a severe sign problem. Thus, the most reliable lattice 
calculations can only be found at small $\mu$ where reasonable algorithms 
are available \cite{deForcrand:2004bk,Lombardo:2004uy}. However, even these 
calculations become difficult especially for large lattices and realistic 
quark masses. Thus, it is fair to say that quantitatively many features of
the $T-\mu$ phase diagram of QCD still remains unclear. A recent review 
of the status of lattice calculations at finite temperature and density can 
be found in \cite{Allton:2005tk,Philipsen:2005mj}.

Given the difficulties of studying the phase diagram of QCD, it is
interesting to consider QCD-like theories which do not suffer from sign 
problems at $\mu\neq 0$ \cite{Kogut:2000ek}.
The sign problem is avoided due to special properties of the fermion action 
which makes the fermion determinant real and positive. As a price,
baryons turn out to be bosons. In spite of this difference QCD-like 
theories provide interesting toy models for QCD. In certain cases they
are also interesting in their own right. A famous example is two-color 
QCD and has been extensively studied over the years both theoretically 
\cite{Kogut:1999iv,Wirstam:1999ds,Ratti:2004ra,Lenaghan:2001sd} 
and numerically \cite{Kogut:1985un,Kogut:1987ev,Kaczmarek:1998wr,Hands:1999md,
Hands:2000ei,Kogut:2001na,Kogut:2003ju,Skullerud:2003yc}. Two-color 
lattice QCD with staggered fermions (2CLQCD) is especially interesting due 
to an enhanced $U(2)$ symmetry at zero quark mass and baryon chemical 
potential. As we will see later, the long distance physics of this theory 
in the $T-\mu$ plane is described by an $O(N)\times O(2)$ ($N=3$ when 
$\mu=0$ and $N=2$ when $\mu \neq 0$) field theory. Such field
theories arise naturally in many condensed matter systems \cite{Kawamura88}
including the studies of spin and charge ordering in cuprate superconductors 
\cite{Zha02} and superfluid transitions in $^3$He \cite{Jones76}. More 
concretely, the normal-to-planar superfluid phase transition in $^3He$ is 
governed by an $O(3)\times O(2)$ field theory \cite{Prato}, which is 
similar to the one that arises in 2CLQCD at $\mu=0$.
Although some progress has been made in uncovering important qualitative 
features of the phase diagram of 2CLQCD, many quantitative questions 
remain: (1) What is the order of the finite temperature chiral transition 
at zero and non-zero chemical potentials? (2) Can the low energy physics at 
small $T$ and $\mu$ be understood quantitatively by an appropriate chiral 
perturbation 
theory \cite{Kogut:2003ju}? (3) What is the order of the phase transition 
that occurs when the lattice gets saturated with baryons at $T=0$? The reason 
for the lack of quantitative progress can be traced to the fact that 
the difficulty of calculations are similar to those in QCD at zero 
chemical potential. Hence, all previous studies have been limited to 
small lattice sizes and relatively large quark masses. Further, diquark
condensation occurs in 2CLQCD and is difficult to study in the 
conventional approach. One usually has to add a diquark source term and
then extrapolate it to zero.

Fortunately, new Monte Carlo algorithms have emerged for lattice gauge
theories in the limit of infinite gauge coupling \cite{Adams:2003cc}
where many of the conventional algorithmic problems disappear. Although 
this strong coupling limit has the worst lattice artifacts, the qualitative 
physics is expected to be preserved. Since most of the current work is 
being done at couplings which can be considered rather strong, the phase 
diagram at these couplings may not be altered significantly as compared 
to the infinite coupling theory. On the other hand, thanks to the 
new algorithms, one can study the chiral limit on large lattices with 
ease at infinite coupling. Thus, studying strong coupling 2CLQCD should 
be a useful step in our general understanding of the subject.
The strong coupling limit attracted the attention of physicists in
the eighties when a variety of qualitative results were obtained
using mean field theory and numerical work \cite{Blairon:1980pk,
Kawamoto:1981hw,Kluberg-Stern:1982bs,Martin:1982tb,Rossi:1984cv,
Wolff:1984we,Dagotto:1986ms,Dagotto:1986gw,Dagotto:1986xt,
Karsch:1988zx,Klatke:1989xy,Boyd:1991fb}. Interestingly, even today
many qualitative questions continue to addressed in this limit
\cite{Bringoltz:2002qc,Bringoltz:2003jf,Bringoltz:2005az}. The 
strong coupling limit of 2CLQCD was originally considered in 
\cite{Dagotto:1986gw,Dagotto:1986xt,Klatke:1989xy} and has been recently 
reviewed in \cite{Nishida:2003uj}. However, many interesting questions, 
including the ones raised above, have remained unanswered even in this 
simplified limit.

In this article we extend the directed path algorithm invented in 
\cite{Adams:2003cc} to study strong coupling 2CLQCD in the chiral limit
and attempt to answer many questions including those raised above. Our 
article is organized as follows. In 
section \ref{model} we discuss the model and the expected physics in 
detail. In section \ref{dimer} we explain the new algorithm which is 
followed by a section in which we discuss the observables and how 
we measure them. Section \ref{results} contains our results which is 
followed by a section where we present a summary of our work along 
with conclusions. A preliminary version of this work appeared in a 
recent conference proceedings \cite{Chandrasekharan:2005dn}.

\section{THE MODEL}
\label{model}

The action of 2CLQCD we study is given by
\begin{equation}
S=-\sum_{x,\alpha}r_\alpha \eta_{\alpha}(x)
\Bigg[
e^{\mu a_t \delta_{t,\alpha}}
\overline{\chi}(x)U_{\alpha}(x)\chi(x+\hat{\alpha})
-e^{-\mu a_t \delta_{t,\alpha}}
\overline{\chi}(x+\hat{\alpha})U_{\alpha}^{\dagger}(x)\chi(x)\Bigg].
\label{scact}
\end{equation}
The Grassmann valued quark fields $\overline{\chi}(x)$ and $\chi(x)$, 
associated to the $3+1$ dimensional lattice site $x$ with 
coordinates $(x_t,x_1,x_2,x_3)$, represent row and column vectors with 
$2$ color components. The components will be denoted as 
$\overline{\chi}_a$ and $\chi_a, a=1,2$. 
The gauge fields $U_{\alpha}(x)$ are elements of $SU(2)$ group and live on 
the links between $x$ and $x+\hat{\alpha}$ where $\alpha=t,1,2,3$. The
factor $r_\alpha=1$ for $\alpha=1,2,3$ and $r_t = 1/a_t$. At weak couplings 
$a_t$ acts as the temporal lattice spacing (assuming spatial lattice spacing 
is $1$). However there is no reason to expect this interpretation to hold at 
strong couplings. Thus, we think of 
$a_t$ as merely an asymmetry factor between spatial and temporal directions. 
It allows us to study the effects of temperature on asymmetric lattices 
and was already used for this purpose in \cite{Boyd:1991fb}. In the 
dimer-baryon loop representation which we will construct later, the 
parameter ${1/(a_t)^2}$ is more natural. By choosing an $L_t\times L^d$ 
lattice (periodic in all directions) we can study thermodynamics in
the $L \rightarrow \infty$ limit at a fixed $L_t$ by defining $T=1/(a_t)^2$ 
as the parameter that represents the temperature. Zero temperature studies
involve the limit $L_t \rightarrow \infty$ with fixed $T$. The parameter 
$\mu$ represents the baryon chemical potential. The absence of the gauge 
action shows that we are in the strong gauge coupling limit.

\subsection{Internal Symmetries}

A detailed discussion of the symmetries of 2CLQCD can be found 
in \cite{Hands:1999md,Nishida:2003uj}. For completeness we review them 
briefly here. We first rewrite eq.~(\ref{scact}) as
\begin{eqnarray}
&&S = -\sum_{x\: {\rm even},\alpha=1,2,3}\eta_{\alpha}(x)
\Bigg[\overline{X_{e}}(x)U_{\alpha}(x)X_{o}(x+\hat{\alpha})-
\overline{X_{e}}(x)U_{\alpha}^{\dagger}(x-\hat{\alpha})X_{o}(x-\hat{\alpha})
\Bigg]
\nonumber \\
&&- \sum_{x\: {\rm even}}\frac{\eta_{t}(x)}{a_t}\Bigg[\overline{X_{e}}(x)
\mathrm{e}^{a_t\mu \sigma_3} U_{t}(x)X_{o}(x+\hat{t})
-\overline{X_{e}}(x)
\mathrm{e}^{-a_t\mu \sigma_3} U_{t}^{\dagger}(x-\hat{t})X_{o}(x-\hat{t})\Bigg]
\nonumber \\
\label{su2act}
\end{eqnarray}
where $\overline{X_{e}}$ and $X_{o}$ are given by
\begin{equation}
\overline{X_{e}}=(\bar{\chi}_{e},-\chi_{e}^{tr}\tau_{2}),
\quad\quad X_{o}=\left(\begin{array}{c}
\chi_{o}\cr -\tau_{2}\bar{\chi}_{o}^{tr}\end{array}\right)
\label{twocompdef}
\end{equation}
In our notation $\vec{\sigma}$ are Pauli matrices that mix $\chi$ and 
$\overline{\chi}^{tr}$ present in $X_o$ and $\overline{X}_e$ while 
$\vec{\tau}$ are Pauli matrices that act on the color space. Thus,
$\tau_{2}U\tau_{2}=U^{\star}$ since $U$ is an element of $SU(2)$. 

Clearly, when $\mu=0$ our model has a $U(2)$ global symmetry:
\begin{equation}
X_{o}\rightarrow VX_{o},\quad
\overline{X}_{e}\rightarrow\overline{X}_{e}V^{\dagger},\qquad 
V = \exp(i\vec{\alpha}\cdot\vec{\sigma}+i\phi) \in U(2).
\label{u2symm}
\end{equation}
This symmetry is reduced to $U_B(1)\times U_C(1)$ in the presence 
of a chemical potential:
\begin{equation}
\begin{array}{ll}
U_B(1): & \quad\quad X_o \rightarrow \exp(i\sigma_3 \phi) X_o, \quad 
\overline{X}_e \rightarrow \overline{X}_e \exp(-i\sigma_3\phi) \cr
U_C(1): & \quad\quad X_o \rightarrow \exp(i\phi) X_o,\quad 
\overline{X}_e \rightarrow \overline{X}_e \exp(-i\phi).
\end{array}
\end{equation}
Here $U_B(1)$ is the baryon number symmetry 
$\chi(x)\rightarrow\mathrm{e}^{i\phi}\chi(x),\quad\overline{\chi}(x)
\rightarrow\overline{\chi}(x)\mathrm{e}^{-i\phi}$ and
$U_C(1)$ is the chiral symmetry of staggered 
fermions $\chi(x)\rightarrow \mathrm{e}^{i\phi\varepsilon(x)}\chi(x)$, 
$\overline{\chi}(x)\rightarrow\overline{\chi}(x)
\mathrm{e}^{i\phi\varepsilon(x)}$ 
where $\varepsilon(x)=(-1)^{x_{t}+x_{1}+x_{2}+x_{3}}$.

\subsection{Properties of the Ground State}

When $\mu=0$ one expects the chiral condensate, which is not 
invariant under the $U(2)$ symmetry, to get a non-zero vacuum expectation 
value. Note that 
\begin{equation}
\hat{\Phi}_i(x) = 
\left\{ \begin{array}{cc}
\frac{-i}{2}X_o^T\sigma_2 [\sigma_i\otimes \tau_2]X_o & x \in \mbox{odd}
\cr
\frac{i}{2}\bar{X}_e [\sigma_i\otimes \tau_2]\sigma_2\bar{X}_e^T & 
x \in \mbox{even}
\end{array}\right.\ \ i=1,2,3 
\label{opara}
\end{equation}
transform as components of a three vector under the $SU(2)$ subgroup of 
the $U(2)$ symmetry group. It is easy to check that
\begin{equation}
\hat{\Phi}_1 = i[\chi_1\chi_2+\bar{\chi_1}\bar{\chi}_2], 
\hat{\Phi}_2 =  [-\chi_1\chi_2+\bar{\chi_1}\bar{\chi}_2], 
\hat{\Phi}_3 = \bar{\chi}\chi = [\bar{\chi_1}\chi_1 + \bar{\chi}_2\chi_2], 
\end{equation}
Thus, the chiral condensate is the third component of a complex three 
vector. In addition all the components carry the same non-zero $U_C(1)$ 
chiral charge. 

The above discussion makes it clear that the chiral condensate being
non-zero is just a matter of choice. More generally, the ground state 
of the theory is such that $\langle \Phi_i \rangle = 
(\hat{n})_i\mathrm{e}^{i\theta}$, where $\hat{n}$ is some constant 
unit three vector.
With this choice, the ground state still remains invariant under a $U(1)$ 
subgroup given by $V=\exp(i\theta \ \hat{n}\cdot\vec{\sigma})$ which
implies that the $U(2)$ symmetry is broken to a $U(1)$ subgroup (note that
the $U(1)$ subgroup must be a part of the $SU(2)$ subgroup of $U(2)$).
When one says that the chiral condensate is non-zero, one implicitly 
chooses $\hat{n}$ along the third direction. This then implies 
$\langle \Phi_1\rangle = \langle \Phi_2\rangle = 0$. However when we 
study the effects of the chemical potential, it is natural to pick the 
ground state such that $\langle \Phi_2 \rangle \neq 0$ and 
$\langle \Phi_1 \rangle = \langle \Phi_3 \rangle = 0$ which implies 
that the diquark condensate, $\langle \chi_1\chi_2\rangle = 
\langle \bar{\chi}_2\bar{\chi}_1\rangle \neq 0$ while the chiral 
condensate vanishes. Note that even though $\langle\bar{\chi}\chi\rangle = 0$ 
the theory still breaks the $U_C(1)$ symmetry since the diquark condensates 
carry a chiral charge. 

When $\mu\neq 0$ the $U(2)$ global symmetry is explicitly broken to 
$U_B(1)\times U_C(1)$. At small $\mu$ both the $U(1)$ symmetries are 
expected to break spontaneously since the diquark condensate continues 
to be non-zero. As $\mu$ increases the density of baryons increases 
and at a critical value $\mu_c$ the lattice becomes saturated with 
baryons which means that for $\mu \geq \mu_c$ the diquark condensate 
vanishes. If this phase transition is second order, the low energy
physics close to $\mu_c$ will be governed by a non-relativistic field 
theory. Renormalization group arguments indicate that this field theory 
is governed by mean field theory in $d\geq 2$ ($d$ represents spatial 
dimensions) \cite{Fisher89}.

\subsection{Finite temperature phase transition}
\label{ftpt}

At high temperatures all the symmetries are expected to be restored. This
implies one must have a finite temperature phase transition for $\mu<\mu_c$.
The order of this transition both at $\mu=0$ and $\mu \neq 0$ in not known.
Since the order parameter at $\mu=0$ is a complex 3-vector, its fluctuations 
are governed by the Landau-Ginzburg (LG) Hamiltonian of the $3$-component 
complex field $\psi(x)$. The $U(2)$ symmetry is manifest as
a $O(3)\times O(2)$ symmetry. Theories with $N$-component complex scalar 
fields with $O(N)\times O(2)$ symmetry are interesting in condensed matter 
physics in describing a variety of materials \cite{Kawamura88}, and are 
described by the action (or classical Hamiltonian),
\begin{equation}
S = \int d^d x \ \Bigg\{
\partial_\mu\psi^*\cdot \partial_\mu\psi
+ r \psi^*\cdot\psi
+ u (\psi^*\cdot \psi)^2 + v |\psi\cdot \psi|^2\Bigg\}.
\label{lg}
\end{equation}
When $r < 0$, depending on the sign of $v$, two classes of ground states 
are allowed (note $u >0$ is necessary for stability). When $v>0$ the 
ground state has a {\em spiral} or {\em helical} 
order, while when $v < 0$ the ground state has {\em collinear} or 
{\em sinusoidal} order. Since we found that $\psi_i \equiv 
\langle \Phi_i \rangle = \hat{n}_i\mathrm{e}^{i\theta}$, we discover that
close to the finite temperature phase transition the long distance physics
of 2CLQCD with massless staggered fermions at $\mu=0$ is described by 
the above complex field theory with $N=3$ and collinear order ($v<0$). 
This field theory is of interest in the study of the normal-to-planar
superfluid transition in $^3$He \cite{Prato}. The question of whether 
the $N=3$ theory with collinear order can be second order still remains 
unresolved. While the $\epsilon$-expansion predicts a fluctuation driven 
first order transition \cite{Kawamura88} recent results claim that a 
second order fixed point indeed exists \cite{Prato}. As we will see later,
in our work we find a weak first order transition. 

It is easy to argue that in the presence of the baryon chemical potential 
the finite temperature phase transition must be governed by a Landau-Ginsburg
Hamiltonian similar to the one above but with $N=2$. Note that the symmetry
in the microscopic theory is reduced to $U(1)\times U(1)$ which is manifest
in the LG theory as an $O(2)\times O(2)$ symmetry. Further, near a finite 
temperature phase transition the presence of a chemical potential does not
usually break charge conjugation symmetries in the low energy effective
theory. For $N=2$ it is possible to define new fields
\begin{equation}
\left( \begin{array}{c} \varphi_1 \cr \varphi_2 \end{array}\right)
= \frac{1}{\sqrt{2}}
\left( \begin{array}{cc} 1 & i \cr i & 1 \end{array}\right)\quad
\left( \begin{array}{c} \psi_1 \cr \psi_2 \end{array}\right),
\end{equation}
such that the LG Hamiltonian becomes
\begin{eqnarray}
S &=& \int d^d x \ \Bigg\{ \Bigg[
\sum_{i=1}^2 \Big(|\partial_\mu\varphi_i|^2 + r |\varphi_i|^2
+ u |\varphi_i|^4\Big)\Bigg]  + 2 (u+2v) |\varphi_1|^2|\varphi_2|^2
\Bigg\}
\label{lg2}
\end{eqnarray}
It is then obvious that when $v<0$ there is always a decoupled $XY$ fixed 
point at $(u+2v)=0$ \cite{Bak76}. Using the knowledge of the critical
exponents in the $XY$ model it can be established that this decoupled
fixed point is stable \cite{Aha76,Pel02}. However, the flow to this fixed 
point is rather slow so that corrections to the $XY$ scaling can be 
substantial until one is very close to the critical point \cite{Vicari}. 
At a non-zero value of $\mu$ we indeed find a second order transition. A 
naive analysis indicates that the critical exponents are different from 
the $XY$ exponents, however when the expected corrections to scaling are 
included the $XY$ exponents can be used to fit our data.

\section{Dimer-Baryonloop Model}
\label{dimer}

\subsection{Dimer-Baryonloop Configurations}

\begin{figure}
\begin{center}
\includegraphics[width=0.5\textwidth]{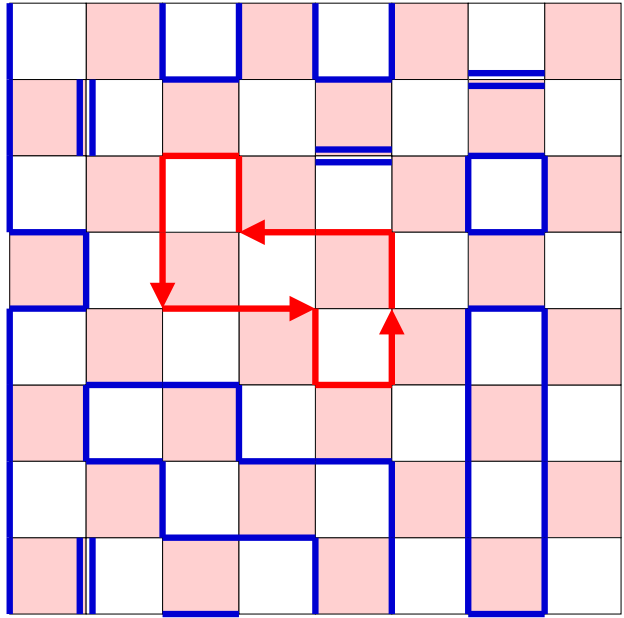}
\end{center}
\caption{\label{fig0} An example of a dimer-baryonloop configuration.}
\end{figure}

One of the computational advantages of the strong coupling limit is that
in this limit it is possible to rewrite the partition function,
\begin{equation}
Z = \int [DU][d\overline\chi d\chi] \exp(-S),
\end{equation}
as a sum over configuration containing gauge invariant objects, namely 
monomers, dimers, and baryonloops 
\cite{Rossi:1984cv,Karsch:1988zx,Dagotto:1986gw,Klatke:1989xy}. Monomers 
are absent in the massless theory. In the case of 2CLQCD a lattice 
configuration $K$ of dimers and baryonloops is constructed as follows: 
(1) Every link of the lattice connecting the site $x$ with the neighboring 
site $x+\hat{\alpha}$ contains either a dimer $k_\alpha(x)=0,1,2$ or a 
directed baryon-bond $b_\alpha(x)=-1,0,1$. When 
$k_\alpha(x)=0$, it means that the link does not contain 
a dimer, while $k_\alpha=1(2)$ implies that the link contains 
a single (double) dimer. Similarly $b_\alpha(x)=0$ means the link does 
not contain a baryon-bond, while $b_\alpha(x) = 1$ means the 
baryon-bond is directed from $x$ to $x+\hat{\alpha}$ and 
$b_\alpha(x)=-1$ means 
it is directed from $x+\hat{\alpha}$ to $x$. We will also allow 
$\hat{\alpha}$ to be negative. Thus, if $\alpha$ was positive, 
$k_{-\alpha}(x)$ and $b_{-\alpha}(x)$ will represent dimers and 
baryon-bonds connecting $x$ with $x-\hat{\alpha}$. (2) If a site is 
connected to
baryon-bonds then it must have exactly one incoming baryon-bond and 
one outgoing baryon-bond. Further it cannot be connected to dimers.
Thus baryon-bonds always form self-avoiding closed baryonloops. 
(3) Every lattice site $x$ that does not contain a baryon-bond must 
satisfy the constraint
\begin{equation}
\sum_\alpha k_\alpha(x) = 2
\end{equation}
where the direction $\alpha$ in the sum takes negative values also. This 
implies that sites connected by single dimers also form a loop which we 
call a dimer-loop. An example of a configuration $K$  is given in 
Figure \ref{fig0}.

\subsection{Updating Algorithms}

Given the set $\{K\}$ of dimer-baryonloop configurations described above,
eq.~(\ref{scact}) can be rewritten as \cite{Klatke:1989xy},
\begin{equation}
Z=\sum_{\{ K\}} \quad
\exp\left( \sum_x \Bigg\{ \Big(k_t(x) + |b_{t}(x)|\Big)\log(T) + 
\frac{2\mu b_{t}(x)}{\sqrt{T}} \Bigg\} \right).
\label{dblpf}
\end{equation}
Since the partition function is written as a sum over positive definite 
terms, a Monte-Carlo algorithm can in principle be designed to the study 
this system. However, the algorithm needs to preserve many constrains. A 
method to do this was developed in \cite{Syl02} in the context of quantum 
spin models and later extended to dimer models in \cite{Adams:2003cc}. We 
will now discuss an extension of these ideas to update the configurations 
$K$. In particular we consider three types of updates: a dimer-baryon loop 
flip update, a dimer update and a baryon update. Below, we discuss each 
of these updates in detail. Remember that we assume 
$k_\alpha(x) = k_{-\alpha}(x+\hat{\alpha})$. The dimer update and baryon 
updates have been described such that this redundant information also 
gets updated automatically.

\subsubsection{Dimer-baryon loop flip update}

This update is based on the observation that every baryon loop's
orientation can be flipped without violating any constraints. Further 
a baryonloop can be converted into a dimer-loop and vice-versa. 
Thus, every loop can be in one of three states: a dimer-loop or a 
baryonloop with two different orientations. Let $Y$ be the subset 
of lattice sites that are connected to either a dimer-loop or a 
baryonloop. $V_Y$ will be the number of these sites. We pick a site from 
$Y$ at random and change the state of the loop ${\cal C}$ 
connected to that site to one of the three allowed states. The 
change can be accomplished using a heat-bath (or similar) update if we 
assign the following weights to the three states: a dimer-loop carries 
a weight $1$ while the baryonloop (in either orientation) carries 
the weight $\exp[2\mu W_t({\cal C})]$, where
\begin{equation}
W_t({\cal C}) = \sum_{x\in {\cal C}} b_t(x)
\end{equation}
Note that $W_t({\cal C})$ changes sign if its orientation is flipped.

\subsubsection{The Dimer Update}

Let $D$ be the set of sites connected by $k_\alpha(x)=1,2$ and $V_D$ be 
the number of such sites. The dimer update changes the configuration on 
a subset of $D$. The update is as follows: 

\begin{enumerate}
\item A lattice site $x\in D$ is selected randomly
\item If the site $x$ lies on a dimer-loop, then there will be two
different directions $\hat{\alpha}$ such that $k_{\alpha}(x)=1$. One 
of these two directions is picked at random. Else there will be one 
direction $\hat{\alpha}$ such that $k_\alpha(x) = 2$. In that case this 
direction is picked. 
\item If the update just started and $x$ is the first site, a virtual 
monomer is created at $x$. If not one is added to $k_{\alpha'}(x)$ where
$\alpha'$ is the direction from which $x$ was reached. One is subtracted 
from $k_\alpha(x)$ and the update moves to the neighboring site 
$y = x+\hat{\alpha}$. We will call $x$ as an ``active
site'' and $y$ as a ``passive site'' in the notation of \cite{Adams:2003cc}.
See next subsection for more details.
\item With all the neighboring sites $y+\hat{\alpha'}$ that belong to 
$D$ a non-zero weight $W_{\alpha'}$ is associated. If $\hat{\alpha'}$ 
is a temporal direction then $W_{\alpha'}=T$, otherwise $W_{\alpha'}=1$. 
If the neighboring site does not belong to $D$ then the weight is zero.
For future reference we define the total dimer weight on the site $y$ 
as $W_D(y)=\sum_{\alpha'} W_{\alpha'}$.
Now based on the weights $W_{\alpha'}$ a heat-bath (or similar) procedure is 
used to pick a new direction $\hat{\alpha'}$. One is subtracted from 
$k_\alpha(y)$ and one is added to $k_{\alpha'}(y)$. The update then moves 
to the neighboring site $x=y+\hat{\alpha'}$.
\item If the site $x$ is not the site which was picked in step
1, the update moves to step 2. Otherwise the site $x$ will contain
one virtual monomer and one direction $\hat{\alpha}$ such that 
$k_\alpha(x)=1$. With probability half the direction $\hat{\alpha}$ is 
picked and the update moves to step 3 and with the remaining probability 
half the virtual monomer on the site is removed and the update ends.
\end{enumerate}

\subsubsection{The Baryon Update}

The third update is just a minor modification of the dimer update.
Let $B$ be the set of sites connected to $k_\alpha(x)=2$ or containing
a baryonloop and $V_B$ the number of such sites.
The baryon update changes the configuration on a
subset of $B$. The update is as follows:

\begin{enumerate}
\item A lattice site $x$ in $B$ is selected randomly.
\item If the site $x$ lies on a baryonloop, then there will be
one direction $\hat{\alpha}$ such that $b_{\alpha}(x)=-1$. This direction
is picked. On the other hand if $x$ is not on a baryonloop then
there will be one direction $\hat{\alpha}$ such that $k_\alpha(x) = 2$. In 
that case this direction is picked. 
\item If the update just started and $x$ is the first site, a virtual 
``diquark'' is created at $x$. Otherwise, one is subtracted from 
$b_{\alpha'}(x)$, where $\alpha'$ is the direction from which $x$
was reached. If $b_{\alpha'}(x) = 0$ after the subtraction then 
$k_{\alpha'}$ is set to 2. One is added to $b_\alpha(x)$ 
and if $k_\alpha(x)=2$ then it is set to zero. The update moves to the 
neighboring site $y = x+\hat{\alpha}$. We will call $x$ as an ``active
site'' and $y$ as a ``passive site'' in the notation of \cite{Adams:2003cc}.
See next subsection for more details.
\item With all the neighboring sites $y+\hat{\alpha'}$ that belong to $B$ 
a non-zero weight $W_{\alpha'}$ is associated. If $\hat{\alpha'}$ is the 
positive temporal direction then $W_{\alpha'}=T\exp(2\mu a_t)$, if it is 
along the negative temporal direction then $W_{\alpha'} = T\exp(-2\mu a_t)$, 
otherwise $W_{\alpha'}=1$. If the neighboring site does not belong to $B$ 
then its weight is zero. For future reference we define the total baryon 
weight on the site $y$ as $W_B(y)=\sum_{\alpha'} W_{\alpha'}$.
Now based on the weights $W_{\alpha'}$ an 
over-relaxation procedure is used to pick a new direction $\hat{\alpha'}$. 
One is subtracted from $b_\alpha(x)$ and if $k_\alpha(x) = 2$ then 
$k_\alpha(x)$ is set to zero. One is added to $b_{\alpha'}(x)$ and if 
$b_{\alpha'}(x)=0$ after the addition then $k_{\alpha'}(x)$ is set to two.  
The update then moves to the site $x=y+\hat{\alpha'}$.
\item If the site $x$ is not the site which was picked in step
1, the update moves to step 2. Otherwise the site $x$ will contain
one virtual diquark and it would have been reached from the direction 
$\hat{\alpha'}$. One is subtracted from $b_{\alpha'}(x)$ and if 
$b_{\alpha'}(x) = 0$ after the subtraction then $k_{\alpha'}$ is set 
to 2. The virtual diquark on the site is removed and the update ends.
\end{enumerate}

\subsection{Active versus Passive Sites}

In the definition of the dimer and baryon updates we have defined active 
and passive sites. The passive sites play an important role during the
measurement of observables. Hence we clarify these two class of sites 
further. Both the dimer and baryon updates are directed loop updates.
They start on a site $x$, which is called an active site. Then they
go through a series of sites which are referred to as passive and active
alternately. Thus the second site is a passive site, the third is an
active site and so on. If the first site is such that $\varepsilon(x) = 1$
then all passive sites $y$, visited during the update, have 
$\varepsilon(y) = -1$ and vice-versa. The weights $W_D(y)$ and $W_B(y)$
for passive sites encountered during the updates will play an special role
in the measurement of correlation functions as discussed below.

\subsection{Detailed Balance and Ergodicity}

Each of the three updates satisfy detailed balance. The proof of detailed 
balance for the the dimer-baryon loop flip update is straight forward and
conventional. On the other hand the proofs for the dimer and baryon updates 
need some understanding of directed loop algorithms. Once this is clear, 
the proof is essentially straight forward. We refer the reader to 
\cite{Adams:2003cc,Syl02} and do not prove detailed balance of these two 
algorithms in this article. The combination of the three updates
makes the algorithm ergodic. To see this we note that there is always 
possible to flip all the baryonloops into dimer-loops. Once this is
done one can use the proof given in \cite{Adams:2003cc} to show that
dimer updates are ergodic in the space of configurations that purely 
consist of dimers.

\section{Observables}

A variety of observables can be measured with our algorithm. We will focus
on the following:

\begin{itemize}

\item[(a)] The chiral two point function is given by
\begin{equation}
G_C(z,z') = \Bigg \langle \bar{\chi}(z)\chi(z) \bar{\chi}(z')\chi(z')
\Bigg\rangle
\end{equation}
and the chiral susceptibility $\chi_{C}$ is
\begin{equation}
\chi_{C}\equiv \frac{1}{V}\sum_{z'} G_C(z,z')
\end{equation}
where $V = L^3 L_t$. Both these observables can be measured easily during 
the dimer update. It is possible to show that \cite{Adams:2003cc}
\begin{equation}
G_C(z,z') = 
\Bigg\langle \sum_y
\frac{V_D\delta_{x,z} \delta_{y,z'}}{W_D(y)}
\Bigg\rangle
\end{equation}
where $x$ is the first site of the dimer update and the sum is over all the 
passive sites $y$ encountered during the update. $V_D$ and $W_D(y)$ were 
defined in the previous section. 

\item[(b)] The diquark two point function is given by
\begin{equation}
G_B(z,z') = 
\Bigg \langle 
\chi_1(z)\chi_2(z) \bar{\chi}_2(z')\bar{\chi}_1(z')
\Bigg\rangle
\end{equation}
and the diquark susceptibility $\chi_{B}$ is given by
\begin{equation}
\chi_{B}\equiv \frac{1}{V}\sum_{z'}  G_B(z,z')
\end{equation}
These can be computed during the baryon update by using
\begin{equation}
G_B(z,z') = 
\Bigg\langle \sum_y
\frac{V_B\delta_{x,z} \delta_{y,z'}}{W_B(y)}
\Bigg\rangle
\end{equation}
where $x$ is the first site picked during the update and the sum is 
over all the passive sites $y$ encountered during the baryon update.
$V_B$ and $W_B(y)$ were defined in the previous section.

\item[(c)] Baryon density $n_{B}$ is defined as
\begin{equation}
n_{B}\equiv\frac{1}{2V}\frac{\partial \ln Z}{\partial\mu}
\end{equation}
and in the dimer-baryon loop flip update it can be measured by the
formula
\begin{equation}
n_{B} = \Bigg\langle 
\frac{V_Y}{V}\frac{W_t({\cal C})}{S({\cal C})}
\frac{\sinh(2\mu W_t({\cal C}))}{1+2\cosh(2\mu W_t({\cal C}))}
\Bigg\rangle
\end{equation}
where ${\cal C}$ is the dimer-baryon loop picked, $V_Y/V$ is the fraction 
of the lattice sites that contain dimer-baryon loops, $S({\cal C})$ is 
the size of the loop and $W_t({\cal C})$(defined earlier) is the temporal 
winding of the baryon loop.

\item[(d)] The helicity modulus associated with the $U(1)$ 
chiral symmetry $Y_{C}$
\begin{equation}
Y_{C} \equiv
\frac{1}{3V_s}\sum_{\mu=1,2,3}
\Bigg\langle\Bigg(\sum_{x}A_{\mu}(x)\Bigg)^{2}\Bigg\rangle
\label{yc}
\end{equation}
where
\begin{equation}
A_{\mu}(x)= \varepsilon(x) \Big( |b_\mu(x)| + k_\mu(x)\Big)
\end{equation}
and $V_s = L^3$. 

\item[(e)] The helicity modulus associated with the $U(1)$ 
baryon number symmetry $Y_{B}$ 
\begin{equation}
Y_{B}\equiv \frac{1}{3V_s}\sum_{\mu=1,2,3}
\Bigg\langle \Bigg(\sum_{x}B_{\mu}(x)\Bigg)^{2}\Bigg\rangle
\label{yb}
\end{equation}
where
\begin{equation}
B_{\mu}(x)= \left[ b_\mu(x) \right]
\end{equation}

\end{itemize}

Both $Y_C$ and $Y_B$ are observables that can be calculated easily
for each configuration $K$ and averaged. Note also that our definitions 
of $Y_C$ and $Y_B$ are natural at finite temperatures.

\section{Results}
\label{results}

\subsection{Zero Chemical Potential}
\label{zcp}

In a finite volume there is no spontaneous symmetry breaking. However, the 
effects of symmetry breaking can still be studied by examining the large 
volume limit of $\chi_C$ and $\chi_B$; if the symmetry is broken in the 
infinite volume limit, we expect these susceptibilities to grow with the 
volume of the system. From symmetry considerations at $\mu=0$ it is 
possible to show that $G_C(z,z') = 2 G_B(z,z')$ which implies 
$\chi_C = 2\chi_B$. Symmetry breaking can also be observed through the 
helicity modulus $Y_C$ and $Y_B$; both must reach a non-zero constant if 
the symmetry breaking pattern is as expected. All these can be understood 
quantitatively using the low energy effective action
\begin{equation}
S_{\rm eff} = \int d^dx \ \Big\{
\frac{B^{2}}{2}(\partial_{\mu}\vec{S})\cdot(\partial_{\mu}\vec{S})
+\frac{F^{2}}{2}(\partial_{\mu}\vec{u})\cdot(\partial_{\mu}\vec{u})
\Big\}
\label{eft}
\end{equation}
where $\vec{S}(x)$ and $\vec{u}(x)$ are unit three and two vector 
fields respectively. Finite size scaling formula for various quantities 
can be obtained following the discussion in \cite{Hasenfratz:1989pk}.
We note that this approach to low energy physics is equivalent to 
others found in the literature \cite{Kogut:1999iv,Kogut:2000ek,Kogut:2003ju}. 

\begin{figure}
\begin{center}
\includegraphics[width=0.8\textwidth]{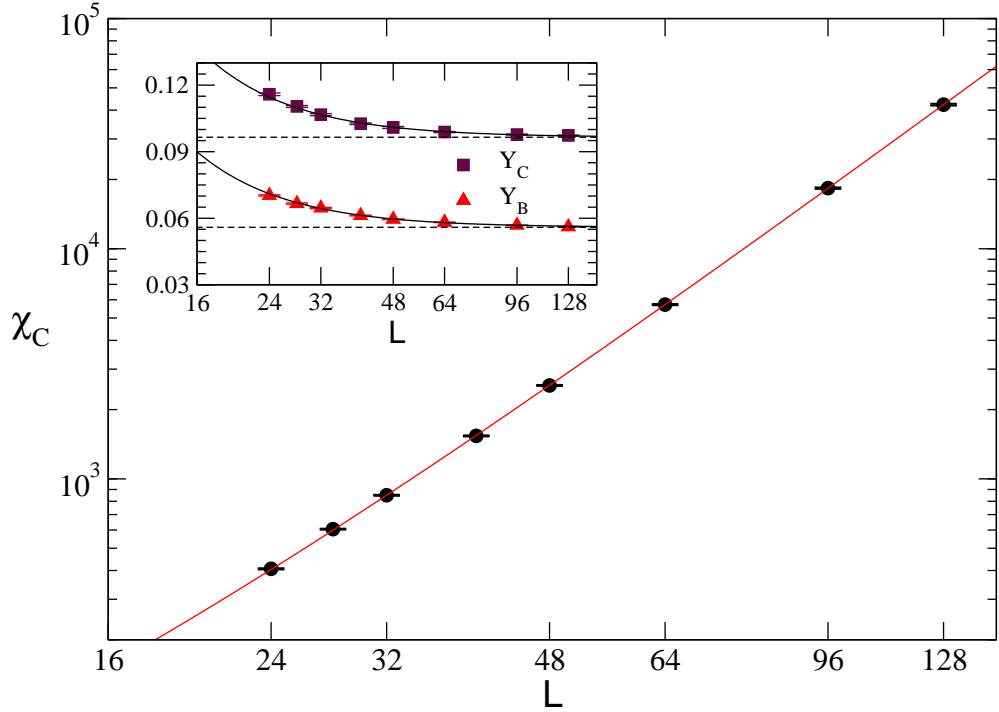}
\end{center}
\caption{\label{fig1} The inset shows the data for $Y_C$ and $Y_B$ as a 
function of $L$. The lines shown are fits to eq.~(\ref{fss2}) which 
yields $F^2=0.0965(5)$ and $B^2=0.0839(6)$. The main figure shows the 
plot of $\chi_C$ versus $L$ for $T=2.918$. The solid line is a fit to 
eq.~(\ref{fss1}) with $F^2$ and $B^2$ fixed to the values quoted above.
The fit yields $\Sigma=0.3364(7)$ and $a=2.6(1)$ with a 
$\chi^2/DOF = 0.5$.}
\end{figure}

At a fixed value of $L_t$ the parameter $T$ can be increased to induce a 
phase transition between a low temperature phase with spontaneous symmetry 
breaking and a high temperature symmetric phase. In order to study this 
phase transition we have performed extensive calculations at a fixed 
$L_t=4$ for different spatial lattice sizes $L$ varying from $16$ to $256$ 
and for many different values of $T$. The low energy effective theory
introduced in eq.~(\ref{eft}) with $d=3$ can then be used to predict the
signatures of the broken phase. We have studied two such signatures: 
(1) $Y_C$ and $Y_B$ go to non-zero constants at large $L$. Extending the 
calculations of \cite{Hasenfratz:1989pk} one can show \cite{Error}
\begin{equation}
Y_C = F^2 + \frac{b}{L} + \frac{c}{L^2}+...; \qquad 
Y_B = \frac{2 B^2}{3} + \frac{b'}{L} + \frac{c'}{L^2}+....
\label{fss2}
\end{equation}
(2) The finite size scaling of the chiral susceptibility is given 
by \cite{Error}
\begin{equation}
\chi_C = \frac{\Sigma^{2}}{6}\Big\{ L^3 + 
\beta_{1}(\frac{2}{B^{2}}+\frac{1}{F^{2}})L^2\Big\} + aL
\label{fss1}
\end{equation}
where $\beta_1 = 0.226$ is the shape coefficient for cubic boxes 
\cite{Hasenfratz:1989pk} and 
$\Sigma/\sqrt{L_t} = \langle \overline{\chi} \chi \rangle$.
Figure \ref{fig1} gives our results at $T=2.918$ which is a value of 
$T$ in the broken phase. The graph shows that the above expectations 
are satisfied well.

\begin{figure}
\begin{center}
\includegraphics[width=0.8\textwidth]{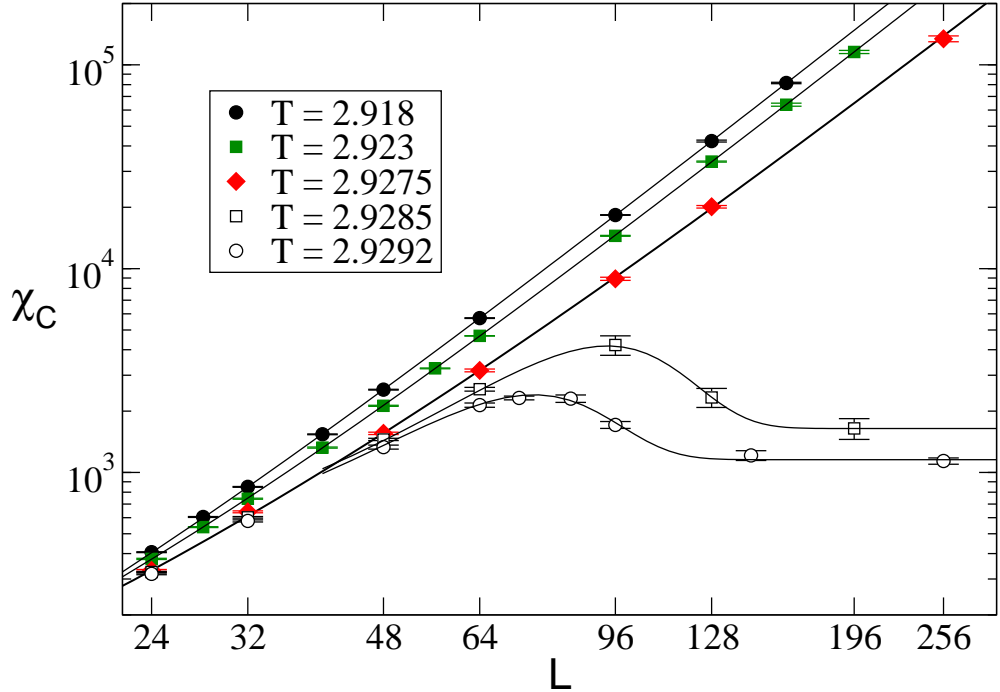}
\end{center}
\caption{\label{fig2} This figure shows the plot of $\chi_C$ versus $L$ for 
different values of $T$ across the phase transition. The solid lines for 
$T\leq 2.9275$ are fits to eq.~(\ref{fss1}) while those for $T=2.9285$ and 
$T=2.9292$ are fits to eq.~(\ref{fss3}). We find 
$\tilde{a}=1650(200),1150(50)$, 
$\tilde{b}=0.030(5),0.027(4)$, $\tilde{c}=2.5(5),2.0(4)$ and 
$\Delta F = 2\times 10^{-6},4\times 10^{-6}$ for the two temperatures. 
All the fits have $\chi^2/DOF$ less than $1$.}
\end{figure}

Figure \ref{fig2} shows the dependence of $\chi_C$ as a function of $L$ for 
different temperatures. Using the data for $L \leq 256$ we find that $\chi_C$ 
increases as $L^3$ for $T \leq 2.9275$, but saturates for $T \geq 2.9285$ as
$L$ becomes large. Thus, we think $T_c$ is between these two temperatures. 
For a second order transition, close to $T_c$, one expects
\begin{subequations}
\begin{eqnarray}
\chi_C L^{\eta-2} &=& [g_0 + g_1(T/T_c-1) L^{1/\nu}+...],
\label{chictc}
\\
Y_C L &=& [f_0 + f_1 (T/T_c-1) L^{1/\nu}+....],
\label{yctc}
\end{eqnarray}
\end{subequations}
which was recently observed in other strong coupling theories
\cite{Chandrasekharan:2003im,Chandrasekharan:2003qv,Chandrasekharan:2004uw}.
Our data does not fit well to this form. Clearly, the data for $\chi_C$ 
shows more structure than can be captured by the above relations. We have 
also verified that $Y_C L$ does not seem to scale as a constant for large 
$L$ anywhere in the range $T=2.928 \pm 0.002$. Hence we think that the 
transition is not second order.  Interestingly, we are able to fit the 
non-monotonic behavior in $\chi_C$ at $T=2.9285$ and $2.9292$ using the 
relation
\begin{equation}
\chi_C = \frac{\tilde{a} + \tilde{b} L^3 
\exp(-\Delta {\cal F} L^3)}{1 + \tilde{c}\exp(-\Delta {\cal F} L^3)}
\label{fss3}
\end{equation}
as long as we use data for $L\geq 48$. This form is natural in the 
presence of two phases (one broken and one symmetric) whose free energy 
densities differ by $\Delta {\cal F}$. This leads us to conclude that 
the phase transition is indeed first order. 

\begin{figure}
\begin{center}
\includegraphics[width=0.8\textwidth]{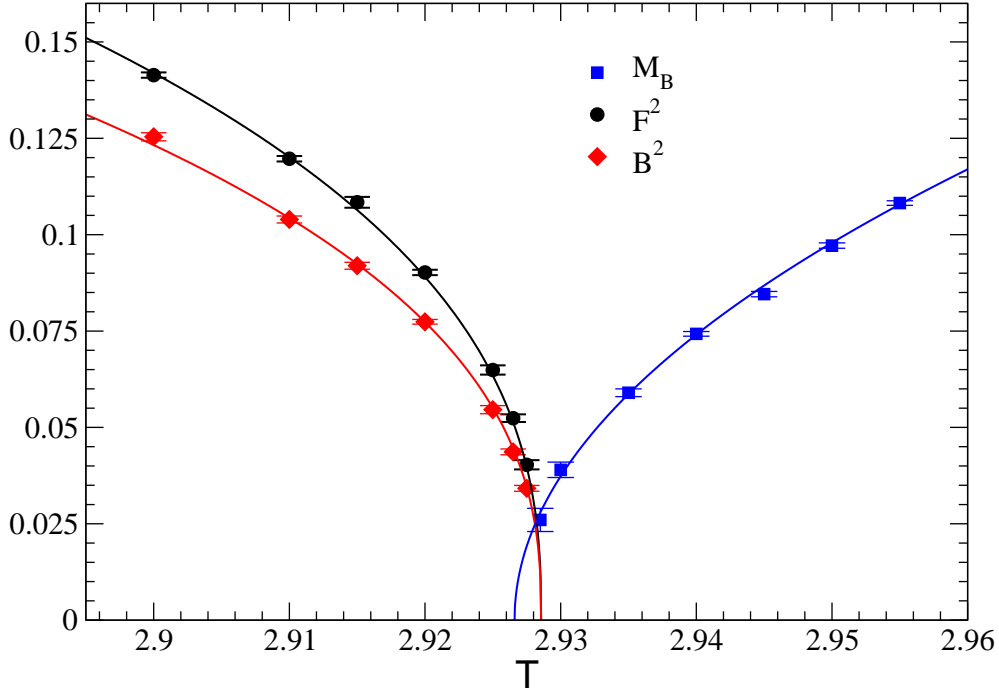}
\end{center}
\caption{\label{fig3} This figure shows the plot of $M_B$, $F^2$ and $B^2$
as a function of $T$. The solid  lines are fits to the data as discussed in
the text.}
\end{figure}

The existence of a first order transition implies that the correlation
lengths at the critical point do not diverge. In that case how big are 
these lengths at the transition? In the high temperature phase one can 
compute screening masses $M_\pi$, $M_B$ from the exponential decay of 
$G_c(z,z')$ and $G_B(z,z')$ respectively, for large spatial separations 
between $z$ and $z'$. At $\mu=0$ we expect $M_\pi=M_B$. In 
the broken phase $F^2$ and $B^2$ have dimensions of mass and are the 
relevant physical scales in the problem. In figure \ref{fig3} we show 
the behavior of $M_B$, $F^2$ and $B^2$ which have been obtained after 
extrapolations to infinite volumes. As can be seen, the correlation 
lengths close to the transition are extremely large, about $40-50$ 
lattice units, indicating that the transition is a weak order transition. 
If the transition was second order, we would have expected 
$F^2 = a_0(T_c-T)^\nu$, $B^2 = b_0(T_c-T)^\nu$ and $M_B = c_0 (T-T_c)^\nu$. 
Interestingly, these relations describe the data reasonably well, but with 
different $T_c$ and $\nu$ in the two phases. In the low temperature phase a 
combined fit of $F^2$ and $B^2$ gives $a_0=0.562(3)$, $b_0=0.488(3)$, 
$T_c=2.92856(4)$ and $\nu=0.387(2)$ with a $\chi^2/DOF=1.5$. On the other 
hand the fit of $M_B$ gives $c_0=0.64(5)$, $T_c=2.9266(6)$ and 
$\nu=0.50(3)$ with a $\chi^2/DOF = 2.2$. A combined fit of all the data 
on both sides of the transition with a single $T_c$ and $\nu$ does not 
fit well confirming our claim that the transition is first order.

\subsection{Non-zero Chemical Potential}

Having established a first order transition at zero chemical potential
we next focus on the finite temperature transition at $\mu=0.3$ with
$L_t=4$. The chemical potential reduces the symmetry of the theory to 
$U_B(1)\times U_C(1)$. At low temperatures 
both the $U(1)$ symmetries are broken leading to two Goldstone bosons. 
The two correlators $G_C(z,z')$ and $G_B(z,z')$ are no longer related: 
$G_c(z,z')$ decays exponentially while $G_B(z,z')$ remains non-zero for 
large $|z-z'|$. This means $\chi_C$ saturates for large $L$ while $\chi_B$ 
grows with the volume and shows the presence of a diquark condensation
and baryon superfluidity. $Y_C$ and $Y_B$ again go to non-zero constants 
at large $L$. In order to determine the finite size scaling formula, we
use the same effective theory as given in eq.~(\ref{eft}) except that now 
both $\vec{S}(x)$ and $\vec{u}(x)$ are unit two vectors. The large $L$ 
limits of $Y_C$ and $Y_B$ are now given by
\begin{equation}
Y_C = F^2 + \frac{b}{L} + \frac{c}{L^2}+...; \qquad 
Y_B = B^2 + \frac{b'}{L} + \frac{c'}{L^2}+....
\label{fss2p}
\end{equation}
The effective field theory also predicts that $\chi_B$ is given by
\begin{equation}
\chi_B = \frac{\Delta^{2}}{2}\Big\{ L^3 + 
\beta_{1}(\frac{1}{F^{2}}+\frac{1}{B^{2}})L^2\Big\} + aL.
\label{fss4}
\end{equation}
where $\Delta/\sqrt{L_t} = \langle\chi_1\chi_2 \rangle 
= \langle\overline{\chi}_2 \overline{\chi}_1 \rangle$.

\begin{figure}
\begin{center}
\includegraphics[width=0.8\textwidth]{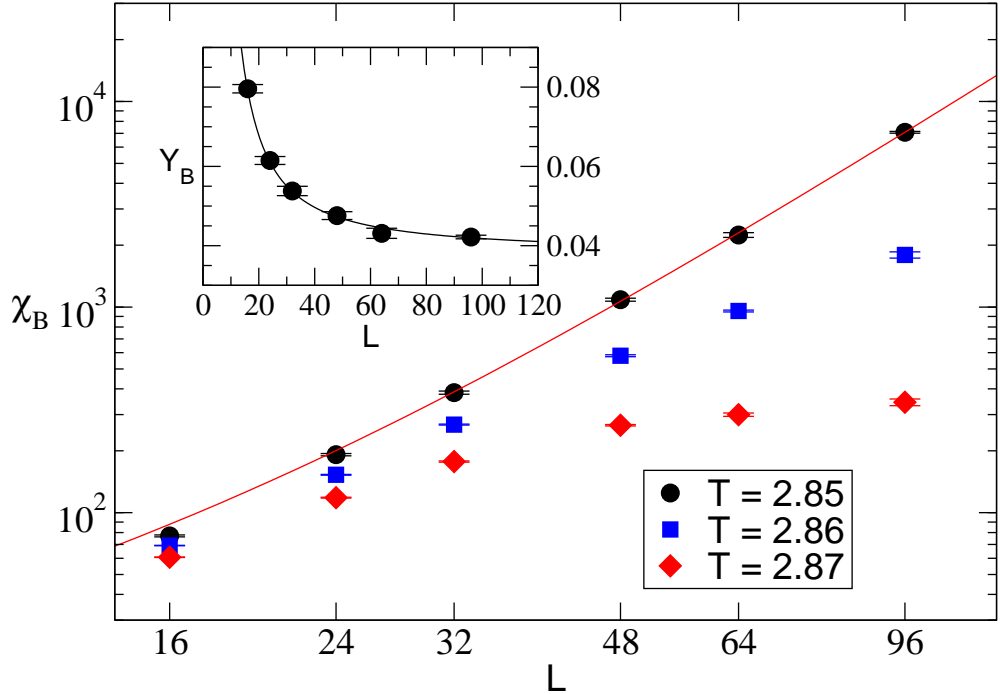}
\end{center}
\caption{\label{fig6} This figure shows $\chi_B$ as a function of $L$ 
at $\mu=0.3$ at three different temperatures close to the critical 
temperature. The inset shows that $Y_B$ goes to a constant as a function
of $L$ at $T=2.85$. The solid line is a plot of eq.~(\ref{fss4}) with
$\Delta=0.117$ and $a=2.4(3)$. Clearly, $\chi_B$ grows as $L^3$ at 
$T=2.85$ but begins to saturate at $T=2.87$ indicating that there is a 
transition between these two temperatures.}
\end{figure}

Interestingly, we find that the values of $Y_B$ and $Y_C$ come
together as we increase the chemical potential, although for 
small temperatures and small chemical potentials we can still 
distinguish between them. On the other hand close to the finite 
temperature phase transition they become indistinguishable. We note 
that the action in eq.~(\ref{lg2}) predicts $F^2=B^2$ close to the 
phase transition. The operator which splits them is irrelevant and 
goes to zero at the transition. However, this explanation does not 
explain why $Y_C$ and $Y_B$ come close to each other as a 
function of $\mu$. This behavior should be examined using effective 
field theory and may emerge naturally, but we have not attempted it 
so far. Figure \ref{fig6} gives our results at $T=2.85$, a value of 
$T$ in the broken phase. 
The inset shows the behavior of $Y_B$ ($Y_C$ looks identical within 
errors). Fitting the data, we find $F^2=B^2=0.0378(11)$ with a 
$\chi^2/DOF$ of around $0.5$. Fixing $F^2$ and $B^2$ to these 
values, our data for $\chi_B$ fits well to eq.~(\ref{fss4}) as long as 
we use $L > 24$. We get $\Delta=0.117(1)$ and $a=2.4(3)$ with
$\chi^2/DOF = 1$. The fit is shown as a solid line in the figure.

\begin{figure}
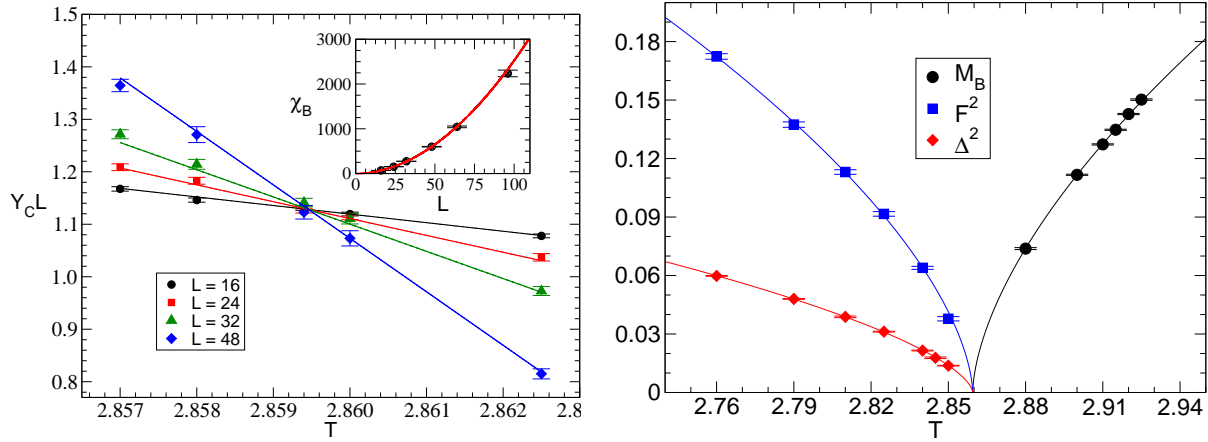

\hbox{
\includegraphics[width=0.48\textwidth]{fig7a.eps}
\includegraphics[width=0.48\textwidth]{fig7b.eps}
}
\caption{\label{fig7} (Left) Plot of $Y_C L$ versus $T$ close to $T_c$ for
various values of $L$. The solid lines are plots of 
$Y_C L = 1.128 - 0.161 (T - 2.85946) L^{1/0.60}$. The inset shows $\chi_B$ 
as a function of $L$ at $T=2.8594$ and the solid line is a plot of 
$0.306 L^{1.958}$. (Right) Plot of $M_B$, $F^2$,$\Delta^2$ versus $T$ 
close to $T_c$. The solid lines show the plots of eq.~(\ref{fdtc}) with 
$a_0=0.70$, $c_0=0.79$,$d_0=0.252$, $\nu=0.610$, $\beta=0.311$ and 
$T_c=2.85946$.}
\end{figure}

Figure \ref{fig6} also shows $\chi_B$ at $T=2.86$ and $2.87$ as a function
of $L$. Unlike the $\mu=0$ case, $\chi_B$ behaves monotonically suggesting 
that the transition could be second order. In order to check this we can
verify if eq.~(\ref{yctc}) is satisfied close to $T_c$. A fit of our data
to this relation gives $T_c=2.85946(7)$, $\nu=0.60(2)$ $f_0=1.128(3)$ and 
$f_1=-0.161(3)$ with a $\chi^2/DOF=1$. In figure \ref{fig7} (plot on the left),
 we show our data and the fit. In the inset of the figure we plot $\chi_B$ at 
$T=2.8594$. Since this value of $T$ is close to $T_c$ we expect 
$\chi_B = g_0 L^{2-\eta}$ 
should describe the data reasonably well. Indeed a fit shows 
$g_0=0.306(8)$, $\eta=0.042(2)$ with a $\chi^2/DOF$ of $1.2$ and is
shown as a solid line in the inset. The values of $F^2$, $\Delta^2$ and 
$M_B$ obtained from the infinite volume extrapolations must also satisfy
\begin{equation}
F^2 = a_0 (T_c-T)^\nu, M_B = c_0 (T-T_c)\nu, \Delta^2 = d_0 (T_c - T)^{2\beta}.
\label{fdtc}
\end{equation}
Figure \ref{fig7} (plot on the right) shows a combined fit of these three 
quantities as a function of $T$ using the value of $T_c$ obtained above. 
We find $\nu=0.610(6)$, $\beta=0.311(5)$, $a_0=0.70(1)$, $c_0=0.79(1)$ and 
$d_0=0.252(3)$ with a $\chi^2/DOF=1.3$. Note that these critical exponents 
do satisfy the hyper-scaling relation $\beta=(d-2+\eta)\nu/2$. Thus,
our data strongly supports the existence of a second order transition.

\begin{figure}
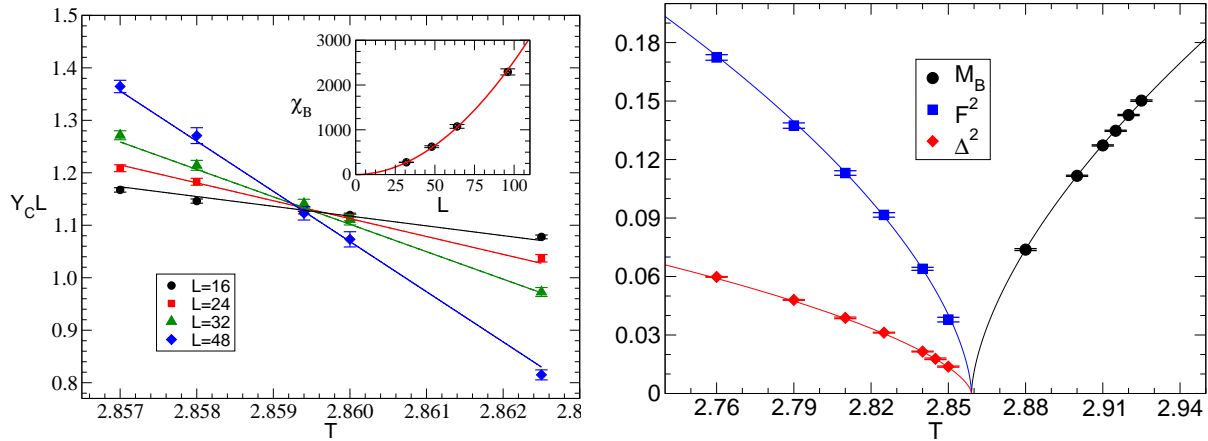

\hbox{
\includegraphics[width=0.48\textwidth]{fig8a.eps}
\includegraphics[width=0.48\textwidth]{fig8b.eps}
}
\caption{\label{fig8} Same as figure \ref{fig7} but now showing the
solid lines show $O(2)$ scaling including corrections. (Left) The 
solid lines represent 
$Y_C L = 2.321 - 1.26 L^{-0.0218} - 0.299 (T - 2.8590) L^{1/0.6715}$. 
The inset shows $\chi_B$ as a function of $L$ at $T=2.8590$ and the 
solid line is a plot of $0.302 L^{1.962}$. (Right) The solid lines 
are plots to eqs.(\ref{fdtcxy}) with $a_0=4.8(5)$, 
$a'_0=-0.8(1)$, $c_0=4.6(4)$, $c'_0=-0.83(8)$, $d_0=2.1(1)$,
$d'_0=-0.89(5)$, $\nu=0.6715$, $\beta=0.3485$, $\omega=0.0218$.}
\end{figure}

Unfortunately, the above results are in contradiction with the expectation 
from section \ref{ftpt}, where it was argued that the critical behavior 
at $\mu \neq 0$ must be governed by three dimensional $XY$ universality 
class. This implies that we should have obtained $\nu=0.6715$, $\beta=0.3485$ 
and $\eta=0.0380$ \cite{Cam01} and not the exponents we found above. As 
discussed in section \ref{ftpt}, the problem is that in a model with 
$U(1)\times U(1)$ symmetry and collinear order, the corrections to the 
$XY$ scaling, due to the irrelevant direction in the $u,v$ plane (see 
section \ref{ftpt}), are rather large. Taking into account the leading 
corrections to scaling one expects 
$Y_C L = f_0 + f'_0 L^{-\omega} + f_1 (T-T_c) L^{1/\nu}$ close 
to $T_c$ where $\omega=0.0218$ \cite{Vicari}. The smallness of $\omega$ 
makes the corrections large for the lattice sizes we have explored.
In fact we find that our data also fits well to this corrected form if 
we use the $XY$ critical exponents and the known value of $\omega$ in the
fits. A combined four parameter fit of our data close to $T_c$ now yields 
$T_c=2.8590(3)$, $f_0=2.321(2)$, $f'_0=-1.260(1)$, $f_1=-0.299(6)$ with
a $\chi^2/DOF=1.6$. This fit is shown in figure \ref{fig8} (the plot on the
left). At $T=2.8590$, 
$\chi_B$ fits well to the form $g_0 L^{2-\eta}$ when $\eta$ is fixed to
$0.0380$. The fit gives $g_0=0.302(4)$ with a $\chi^2/DOF$ of $0.57$
(solid line in the inset of figure \ref{fig8}). When the scaling corrections 
are included in $F^2$, $\Delta^2$ and $M_B$ one gets 
\begin{subequations}
\label{fdtcxy}
\begin{eqnarray}
F &=& a_0 (T_c-T)^\nu(1 + a'_0(T_c-T)^{\omega\nu}),\qquad 
\\
M_B &=& c_0 (T-T_c)^\nu(1+c'_0(T-T_c)^{\omega\nu}), \qquad
\\
\Delta^2 &=& d_0 (T_c - T)^{2\beta}(1+d'_0(T_c-T)^{\omega\nu}).
\end{eqnarray}
\end{subequations}
Fixing $T_c=2.859$ and using the $XY$ critical exponents and $\omega$ as
above, a combined fit again works very well and is shown in the plot on
the right in figure \ref{fig8}. We get $a_0=4.8(5)$, $a'_0=-0.8(1)$, 
$c_0=4.6(4)$, $c'_0=-0.83(8)$, $d_0=2.1(1)$ and $d'_0=-0.89(5)$ with a 
$\chi^2/DOF=1$.

\begin{figure}[t]
\begin{center}
\includegraphics[width=0.8\textwidth]{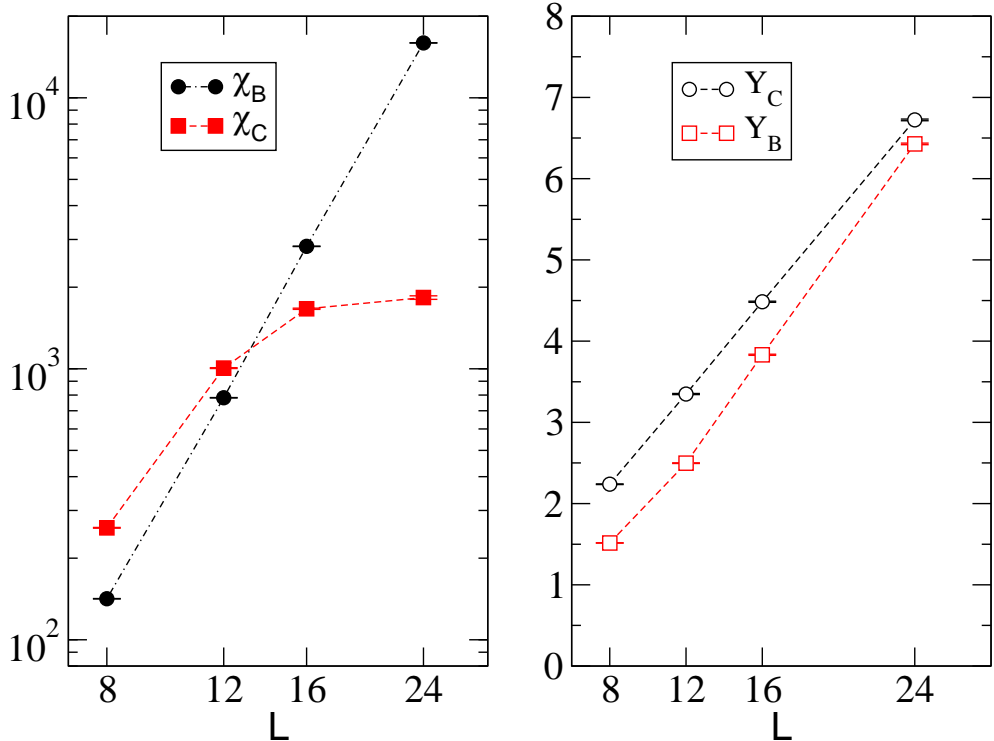}
\end{center}
\caption{\label{fig4} Results at $T=1.0$ and $\mu=0.01$. Note that $\chi_B$ 
grows with the volume while $\chi_C$ saturates. Both $Y_C$ and $Y_B$ grow 
linearly with $L$. The lines are drawn to guide the eye.}
\end{figure}

\begin{figure}[t]
\begin{center}
\includegraphics[width=0.8\textwidth]{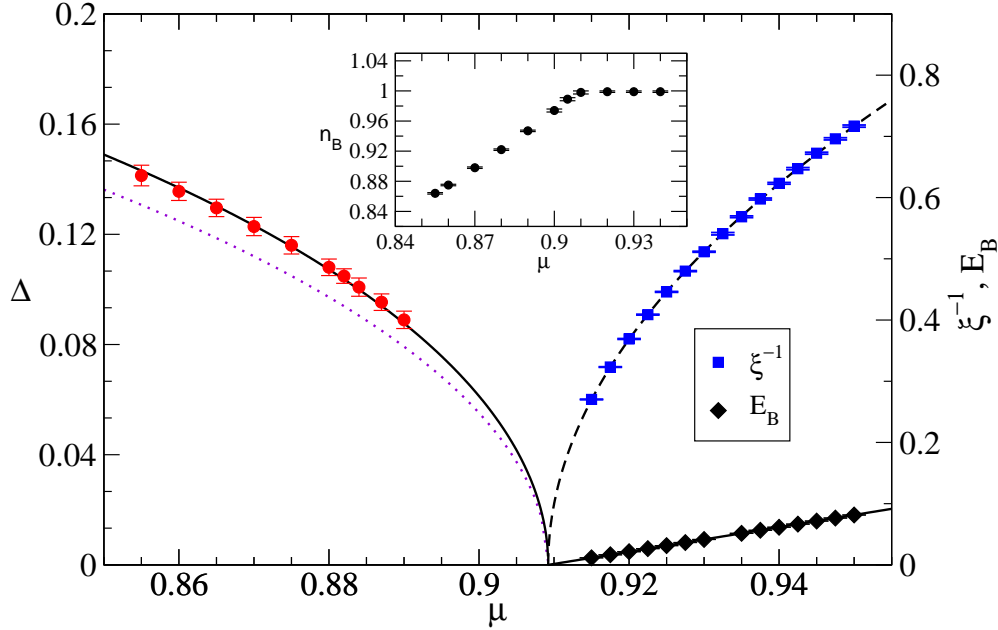}
\end{center}
\caption{\label{fig5} The figure shows our results for $\Delta$, $\xi^{-1}$ 
and $E_B$ as a function of $\mu$. For $\mu < \mu_c$ the dotted line is the 
mean field result for $\Delta$ given in eq.~(\ref{mfdelta}) and the solid 
line contains the one loop corrections. For $\mu > \mu_c$ the dashed line 
is the plot of $3.549 \sqrt{\mu-\mu_c}$ and the solid line is the
plot of $2 (\mu-\mu_c)$. We have used $\mu_c=0.90922$ here. The inset 
shows the baryon density $n_B$ as a function of $\mu$.}
\end{figure}

\subsection{Zero Temperature}

Next we turn to the physics at zero temperature. For this purpose we compute 
quantities with $L_t=L$ at $T=1.0$ for various values of $\mu$ and $L$. We
now expect
\begin{equation}
\chi_B \sim \frac{\Delta^2}{2} L^4
\end{equation}
where $\Delta = \langle \chi_1\chi_2 \rangle = 
\langle \bar{\chi}_2 \bar{\chi}_1\rangle \neq 0$. Note that since we use
the same finite size scaling form in four and three dimensions, $\Delta$ 
is normalized differently here as compared to the finite temperature case. 
The chiral susceptibility $\chi_C$, must again saturate with $L$ at any 
non-zero value of $\mu$. Finally, due to our definitions (see eqs.(\ref{yc}),
(\ref{yb})) both the helicity modulus $Y_C$ and $Y_B$, grow linearly with 
$L$ for large $L$. These expectations emerge nicely in our calculations
as can be seen in figure \ref{fig4}.

As the chemical potential increases the average number of baryons in the 
ground state increases. At some critical chemical potential $\mu_c$, the 
ground state has a baryon on every lattice site. Since the baryons behave 
as hard-core bosons, at $\mu_c$ there must be a phase transition to a 
phase where superfluidity is no longer present. We now focus on this phase 
transition. Renormalization group arguments show that this phase transition 
must be governed by mean field theory \cite{Fisher89}. A mean field analysis 
was performed recently in \cite{Nishida:2003uj} and the critical chemical 
potential was found to be $\mu_c = 0.5\cosh^{-1}(\sqrt{10}) = 0.909223..$. 
The diquark condensate was shown to be \cite{Comm1}
\begin{equation}
\Delta = \sqrt{\frac{1}{18}\left(\sqrt{10}-\cosh(2\mu)\right)}
\label{mfdelta}
\end{equation}
In order to check these results we have computed the diquark condensate
$\Delta$ by fitting our $\chi_B$ data to the relation
$\chi_B = \frac{\Delta^2}{2}[L^4 + a'' L^2 + b'']$. Figure \ref{fig5} 
shows our data along with the mean field result \cite{Nishida:2003uj} 
and the result with one-loop corrections \cite{Jia06}. Clearly, 
the one-loop corrections are necessary before connection with mean field 
theory can be established.

The fact that this phase transition is driven due to the saturation of 
the lattice with baryons can be seen in the inset of figure \ref{fig5}, 
where the baryon density is plotted as a function of $\mu$. For $\mu >\mu_c$ 
it costs an energy $E_B$ to remove a baryon. This energy gap grows linearly 
as $(\mu-\mu_c)$. Thus for $\mu > \mu_c$ one obtains a phase containing
non-relativistic particles whose dispersion relation for small momenta
looks like $E(p) = E_B + p^2/2 M_k$ which leads to a spatial correlation 
length $\xi = 1/\sqrt{2 E_B M_k}$. Since we expect $\xi$ to scale as 
$1/\sqrt{(\mu-\mu_c)}$ close to $\mu_c$ one expects the kinetic mass 
$M_k$ to be a constant. Figure \ref{fig5} shows the plot of $E_g$ and 
$\xi^{-1}$ 
as a function of $\mu$. We find $\xi^{-1} = 3.549(4) \sqrt{(\mu- 0.90920(3))}$ 
with a $\chi^2/DOF=0.26$ and $E_B= 2.000(2) (\mu-0.90922(1))$ with a 
$\chi^2/DOF=0.35$. Again $\mu_c$ is in excellent agreement with the mean 
field result which provides strong evidence that the phase transition is 
second order and belongs to the mean field universality class. From the 
behavior of $E_B$ and $\xi$ we find $M_k = 3.12(5)$. 

\section{Discussion and Conclusions}

\begin{figure}
\begin{center}
\includegraphics[width=0.5\textwidth]{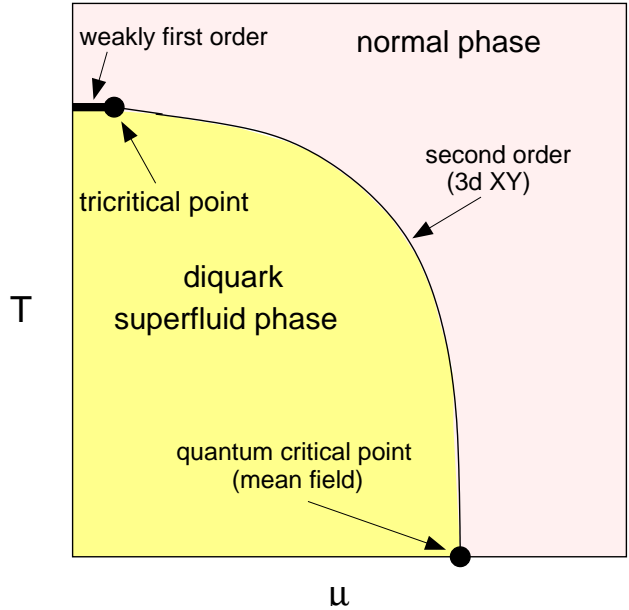}
\end{center}
\caption{\label{pdiag} The phase diagram of two-color lattice QCD 
with staggered fermions at strong couplings.}
\end{figure}

In this work we constructed an efficient cluster algorithm and 
studied the phase structure of two color lattice QCD with massless 
staggered fermions in the strong coupling limit. We found that the 
finite temperature phase transition at zero chemical potential is 
weakly first order, while the same transition at an intermediate 
value of the chemical potential was second order. This second order 
transition was found to be in the universality class of the three 
dimensional $XY$ model as expected from theoretical arguments. However, 
in order to show this we needed to include the large corrections 
to scaling expected in the theory. The quantum phase transition at 
zero temperature between a baryon superfluid phase and a normal phase 
was also found to be second order in the mean field universality 
class. The physics in the normal phase was that of interacting 
non-relativistic particles. Based on these observations, we can
attempt to draw the full diagram of two color QCD in the strong 
coupling limit. Our proposal is shown in figure \ref{pdiag}. 

It is important to understand how this phase diagram will change as 
we go to weaker couplings and towards the continuum limit. Clearly, 
when the four flavor nature of staggered fermions becomes important, 
it can begin to change significantly. However, for couplings that are
currently explored, universality arguments suggest that the phase 
diagram will qualitatively remain the same although quantitatively
the values of the critical points will mostlikely get affected. It 
is also interesting to ask why the transition at $\mu=0$ is so weak. 
As discussed in section \ref{model}, 
renormalization group studies based on the $\epsilon$-expansion 
indicate that the transition will generically be a fluctuation driven 
first order transition \cite{Kawamura88}. Such a transition could be weak 
as we observe. On the other hand, recent work based on high order 
perturbation theory combined with resummation techniques do not rule 
out the possibility of a second order transition \cite{Prato}. This 
means we are in the vicinity of a tricritical point and by changing
some parameter in the theory we could change the transition to 
second order. When this occurs, it is likely that the weakly first
order line in the above phase diagram will disappear. Such a scenario 
can in principle be investigated by introducing more tunable parameters 
within our model and by studying their effects on the order of the 
transition.

It is amusing to note that the above phase diagram is different
from the standard phase diagram in QCD where the baryon chemical 
potential induces a first order transition instead of weakening it. 
This non-standard scenario was discussed in \cite{Philipsen:2005mj} 
as a possibility in QCD and has also been seen in the potts model
simulations \cite{Kim:2005ck}. Finally, we note that the physics 
close to the quantum critical point is also interesting from the 
point of view of non-relativistic field theory. Since the model 
has a $U(1)\times U(1)$ symmetry, the low energy effective field 
theory here is richer than in a theory with a simple $U(1)$ 
particle number symmetry.

\section*{Acknowledgments}

We would like to thank E. Vicari for his time for explaining to us
the existence of the decoupled $XY$ fixed point and how the corrections
to scaling may be important in our work. We also thank Ph. de Forcrand,
S. Hands, C. Strouthos, T. Mehen, R. Springer, D. Toublan and U.-J.Wiese 
for many helpful comments. This work was supported in part by the Department 
of Energy grant DE-FG02-03ER41241. The computations were performed on 
the CHAMP, a computer cluster funded in part by the DOE.

\newpage

\begin{center}
{\bf Appendix A: Exact results on a $2 \times 2$ Lattice}
\end{center}

Below we list the exact expressions for various observables on a 
$2\times 2$ lattice. We have tested the algorithm against these
exact results. Tables \ref{tabe1} and \ref{tabe2} show the 
comparison of exact results against those obtained using the 
algorithm.

\begin{equation}
\chi_{C}=2\frac{1}{4}
\frac{8T^{3}(3+2\cosh(\frac{4\mu}{\sqrt{T}}))
+32(T^{2}+T)+40}{9T^{4}+2T^{4}
(1+\cosh(\frac{8\mu}{\sqrt{T}})+
6\cosh(\frac{4\mu}{\sqrt{T}}))+25+16T^{2}
(2+\cosh(\frac{4\mu}{\sqrt{T}}))}
\end{equation}

\begin{equation}
\chi_{B}=\frac{1}{8}\frac{(48T^{3}+32T^{3}\cosh(\frac{4\mu}{\sqrt{T}})+64T)\cosh(\frac{2\mu}{\sqrt{T}})+8(1+\cosh(\frac{4\mu}{\sqrt{T}}))4T^{2}+80}{9T^{4}+2T^{4}(1+\cosh(\frac{8\mu}{\sqrt{T}})+6\cosh(\frac{4\mu}{\sqrt{T}}))+25+16T^{2}(2+\cosh(\frac{4\mu}{\sqrt{T}}))}\label{37}\end{equation}

\begin{equation}
n_{B}=\frac{1}{4}\frac{8T^{4}\sinh(\frac{8\mu}{\sqrt{T}})+16(2T^{2}+1.5T^{4})\sinh(\frac{4\mu}{\sqrt{T}})}{9T^{4}+2T^{4}(1+\cosh(\frac{8\mu}{\sqrt{T}})+6\cosh(\frac{4\mu}{\sqrt{T}}))+25+16T^{2}(2+\cosh(\frac{4\mu}{\sqrt{T}}))}\label{38}\end{equation}

\begin{equation}
Y_B = \frac{1}{4}\frac{32T^{2}\cosh(\frac{4\mu}{\sqrt{T}}))+32T^{2}+80}{9T^{4}+2T^{4}(1+\cosh(\frac{8\mu}{\sqrt{T}})+6\cosh(\frac{4\mu}{\sqrt{T}}))+25+16T^{2}(2+\cosh(\frac{4\mu}{\sqrt{T}}))}\label{39}\end{equation}

\begin{equation}
Y_C = \frac{1}{4}\frac{32T^{2}\cosh(\frac{4\mu}{\sqrt{T}}))+64T^{2}+80}{9T^{4}+2T^{4}(1+\cosh(\frac{8\mu}{\sqrt{T}})+6\cosh(\frac{4\mu}{\sqrt{T}}))+25+16T^{2}(2+\cosh(\frac{4\mu}{\sqrt{T}}))}\label{40}\end{equation}

\begin{center}

\begin{table}[ht]
\begin{tabular}{|c|c|c|}
\hline 
$\mu$& \multicolumn{2}{c|}{$\chi_C$} \tabularnewline 
\hline
& exact & algorithm \tabularnewline
\hline \hline 
0.0& 0.734694& 0.73463(33) \tabularnewline
\hline 
0.1& 0.719686& 0.71926(27) \tabularnewline
\hline 
0.5& 0.412785& 0.41267(31) \tabularnewline
\hline 
2.0& 0.001343& 0.00138(3)\tabularnewline
\hline
\end{tabular}
\begin{tabular}{|c|c|}
\hline 
\multicolumn{2}{|c|}{$\chi_B$} \tabularnewline 
\hline
exact& algorithm\tabularnewline
\hline \hline 
0.367347& 0.36740(6)\tabularnewline
\hline 
0.366698& 0.36674(6)\tabularnewline
\hline 
0.324066& 0.32405(7)\tabularnewline
\hline 
0.018948& 0.01895(1)\tabularnewline
\hline
\end{tabular}
\begin{tabular}{|c|c|}
\hline 
\multicolumn{2}{|c|}{$n_B$} \tabularnewline 
\hline
exact& algorithm\tabularnewline
\hline \hline 
0.0& 0.0 \tabularnewline
\hline 
0.074563& 0.07441(8)\tabularnewline
\hline 
0.462172& 0.46224(20)\tabularnewline
\hline 
0.997655& 0.99764(3)\tabularnewline
\hline
\end{tabular}

\begin{tabular}{|c|c|c|}
\hline 
$\mu$ & \multicolumn{2}{|c|}{$Y_B$} \tabularnewline 
\hline
& exact& algorithm\tabularnewline
\hline \hline 
0.0 & 0.367347& 0.36740(26)\tabularnewline
\hline 
0.1 & 0.363055& 0.36347(21)\tabularnewline
\hline 
0.5 & 0.254861& 0.25487(20)\tabularnewline
\hline 
2.0 & 0.001339& 0.00135(2)\tabularnewline
\hline
\end{tabular}
\begin{tabular}{|c|c|}
\hline 
\multicolumn{2}{|c|}{$Y_C$} \tabularnewline 
\hline
exact& algorithm\tabularnewline
\hline\hline 
0.448980& 0.44887(34)\tabularnewline
\hline 
0.442306& 0.44258(32)\tabularnewline
\hline 
0.289955& 0.28981(24)\tabularnewline
\hline 
0.001340& 0.001337(13)\tabularnewline
\hline
\end{tabular}
\caption{\label{tabe1} Exact versus Monte Carlo results at $T=1$.}
\end{table}

\begin{table}[ht]
\begin{tabular}{|c|c|c|}
\hline 
$\mu$ & \multicolumn{2}{|c|}{$\chi_C$} \tabularnewline 
\hline
& exact& algorithm\tabularnewline
\hline \hline 
0.0& 0.302981& 0.30305(11)\tabularnewline
\hline 
0.1& 0.299581& 0.29958(11)\tabularnewline
\hline 
0.5& 0.229674& 0.22987(12)\tabularnewline
\hline 
2.0& 0.012903& 0.01295(5)\tabularnewline
\hline
\end{tabular}
\begin{tabular}{|c|c|}
\hline 
\multicolumn{2}{|c|}{$\chi_B$} \tabularnewline 
\hline
exact& algorithm\tabularnewline
\hline \hline 
0.151491& 0.15153(10)\tabularnewline
\hline 
0.150955& 0.15095(4)\tabularnewline
\hline 
0.137719& 0.13763(2)\tabularnewline
\hline 
0.034431& 0.03444(1)\tabularnewline
\hline
\end{tabular}
\begin{tabular}{|c|c|}
\hline 
\multicolumn{2}{|c|}{$n_B$} \tabularnewline 
\hline
exact& algorithm\tabularnewline
\hline \hline 
0.0& 0.0 \tabularnewline
\hline 
0.082043& 0.08218(13)\tabularnewline
\hline 
0.403837& 0.40373(17)\tabularnewline
\hline 
0.966865& 0.96678(9)\tabularnewline
\hline
\end{tabular}

\begin{tabular}{|c|c|c|}
\hline 
$\mu$& \multicolumn{2}{|c|}{$Y_B$} \tabularnewline 
\hline
& exact& algorithm\tabularnewline
\hline \hline 
0.0&0.066076& 0.066028(55)\tabularnewline
\hline 
0.1& 0.065598& 0.065625(56)\tabularnewline
\hline 
0.5& 0.054750& 0.054770(66)\tabularnewline
\hline 
2.0& 0.004203& 0.004195(13)\tabularnewline
\hline
\end{tabular}
\begin{tabular}{|c|c|}
\hline 
\multicolumn{2}{|c|}{$Y_C$} \tabularnewline
\hline 
exact& algorithm\tabularnewline
\hline \hline 
0.095085& 0.095026(87)\tabularnewline
\hline 
0.094062& 0.094109(66)\tabularnewline
\hline 
0.072868& 0.072927(78)\tabularnewline
\hline 
0.004284& 0.004277(13)\tabularnewline
\hline
\end{tabular}
\caption{\label{tabe2} Exact versus Monte Carlo results at $T=3$.}
\end{table}
\end{center}

\end{document}